# Regenerative oscillation and four-wave mixing in graphene optoelectronics

T. Gu[1*], N. Petrone[1], J. F. McMillan[1], A. van der Zande[1], M. Yu[2], G. Q. Lo[2], D. L. Kwong[2], J. Hone[1], and C. W. Wong[1*]

[1] Columbia University, New York, NY 10027, USA.

[2] The Institute of Microelectronics, 11 Science Park Road, Singapore Science Park II Singapore 117685, Singapore.

Email: * tg2342@columbia.edu and cww2104@columbia.edu

**The unique linear and massless band structure of graphene, in a purely two-dimensional Dirac fermionic structure, have led to intense research spanning from condensed matter physics [1-5] to nanoscale device applications covering the electrical [6-7], thermal [8-9], mechanical [10] and optical [11, 12] domains. Here we report three consecutive first-observations in graphene-silicon hybrid optoelectronic devices: (1) ultralow power resonant optical bistability; (2) self-induced regenerative oscillations; and (3) coherent four-wave mixing, all at a few femtojoule cavity recirculating energies. These observations, in comparison with control measurements on solely monolithic silicon cavities, are enabled only by the dramatically-large and ultrafast $\chi^{(3)}$ nonlinearities in graphene and the large *Q/V* ratios in wavelength-localized photonic crystal cavities. These third-order nonlinear results demonstrate the feasibility and versatility of hybrid two-dimensional graphene-silicon nanophotonic devices for next-generation chip-scale high-speed optical communications, radio-frequency optoelectronics, and all-optical signal processing.**

Sub-wavelength nanostructures in monolithic material platforms have witnessed rapid advances towards chip-scale optoelectronic modulators [13-16], photoreceivers [17-18], and high-bitrate signal processing architectures [19-20]. Coupled with ultrafast nonlinearities as a new parameter space for optical physics [21], breakthroughs such as resonant four-wave mixing [22] and parametric femtosecond pulse characterization [23-24] have been described. Recently, graphene – with its broadband dispersionless nature and large carrier mobility – has been examined for its gate-variable optical transitions [25-26] towards broadband electroabsorption modulators [27] and photoreceivers [28-29] including planar microcavity-enhanced photodetectors [30-31], as well as saturable absorption for mode-locking [32]. Due to its linear band structure allowing interband optical transitions at all photon energies, graphene has been suggested as a material with large $\chi^{(3)}$ nonlinearities [33]. In this Letter we demonstrate the exceptionally high third-order nonlinear response of graphene with a wavelength-scale localized photonic crystal cavity, enabling ultralow power optical bistable switching, self-induced regenerative oscillations, and coherent four-wave mixing at femtojoule cavity energies on the semiconductor chip platform. The structure examined is a hybrid graphene-silicon cavity (as illustrated in Figure 1), achieved by rigorous transfer of monolayer large-area graphene sheet onto air-bridged silicon photonic crystal nanomembranes with minimal linear absorption and optimized optical input/output coupling. This optoelectronics demonstration is complemented with recent examinations of large-area [34-35] graphene field-effect transistors and analog circuit designs [36] for potential large-scale silicon integration.

Figure 1 illustrates the graphene-cladded photonic crystal nanomembranes investigated. The optical nanocavity is a point-defect photonic crystal *L3* cavity (with three missing holes) [35-36], with nearest holes at the cavity edges tuned by 0.15*a* (where *a* is the photonic crystal lattice constant). The *L3* cavity is side coupled to a photonic crystal line defect waveguide for optical transmission measurements. Chemical vapor deposition (CVD) grown graphene is wet-transferred onto the silicon nanomembrane [39-40] (see Methods; and Supplementary Information, Section S1), with the graphene heavily *p*-doped, on a large sheet without requiring precise alignment. As shown in Figure 1b, the single layer graphene is identified by Raman spectroscopy via the full-width half-maximum of the *G* and *2D* band peaks (34.9 cm$^{-1}$ and 49.6 cm$^{-1}$ respectively) and the *G*-to-*2D* peak intensity ratio of ~ 1 to 1.5. The *G* band lineshape is a single and symmetrical Lorentzian indicating good uniformity graphene [41]. Heavily doped graphene is specifically prepared to achieve optical transparency in the infrared with negligible linear losses, as the Fermi level is below the one-photon interband optical transition threshold [27] (Figure 1c inset) and intraband graphene absorption is near-absent in the infrared [42].

Transverse-electric (TE) polarization laser light is launched onto the optical cavity and evanescently coupled to the monolayer graphene. As shown in Figure 1d, the cavity transmission spectra, performed with tunable continuous-wave laser sources at 0.6 mW, shows a consistent and large resonance red-shift of 1.2 nm/mW, approximately 4× larger than that of our near-identical control cavity without graphene (more measurements detailed in the Supplementary Information, Section S3). The low power "cold cavity" transmissions taken at 2.5 µW input powers depict intrinsic *Q*s of 23,000 and loaded *Q*s of 7,500, with background Fabry-Perot oscillations arising from the input/output facet coupling reflections (~ 0.12 reflectivity). The high power cavity transmission is not only red-shifted to *outside* the cold cavity lineshape full-width base but also exhibit a Fano-like asymmetric lineshape, with good matching to our coupled-mode model predictions (Supplementary Information, Section S3). We also note that with the transferred monolayer graphene onto only the short photonic crystal regions the total fiber-chip-fiber transmission is decreased by less than 1 dB, slightly better than the 5-dB additional loss in recent modified graphene-fiber linear polarizers [43] (with different cavity or propagation lengths and evanescent core coupling). We emphasize that, for the same increased cavity power on a monolithic silicon cavity without graphene, both the control experiment and numerical models show a negligble thermal red-shift of 0.1 nm/mW, for the power levels and the specific loaded cavity $Q^2/V$ values [of $4.3 \times 10^7 (\lambda/n)^3$] investigated here.

The large frequency shifts from the graphene-cladded hybrid photonic cavity is next investigated for low-threshold optical bistability. Figure 2a shows the observed bistability at 100 µW threshold powers for a loaded cavity $Q$ of 7,500 ($Q_{\text{intrinsic}}$ of 23,000), with cavity – input laser detuning $\delta$ of 1.5 [with $\delta$ defined as $(\lambda_{laser} - \lambda_{cavity})/\Delta\lambda_{cavity}$, where $\Delta\lambda_{cavity}$ is the cold cavity full-width half-maximum linewidth]. The steady-state bistable hystersis at a detuning of 1.7 is also illustrated in Figure 2a. The dashed lines show the coupled-mode theory numerical predictions of the hybrid cavity, including first-order estimates of the graphene-modified thermal, linear and nonlinear loss, and free carrier parameters (detailed in the Supplementary Information, Sections S2 and S3). We also note the heavily-doped graphene has a two-photon absorption at least several times larger than silicon, described by its isotropic bands for interband optical transitions [43], leading to increased free carrier densities/absorption and overall enhanced thermal red-shift.

To verify the bistable switching dynamics, we input time-varying intensities to the graphene-cladded cavity, allowing a combined cavity power – detuning sweep. Figure 2b shows an example time-domain output transmission for two different initial detunings [$\delta_{(t=0)} = -1.3$ and $\delta_{(t=0)} = 1.6$] and for an illustrative triangular-waveform drive, with nanosecond resolution on an amplified photoreceiver. With the drive period at 77 ns, the observed thermal relaxation time is ~

40 ns. Cavity resonance dips (with modulation depths ~ 3-dB in this example) are observed for both positive detuning (up to 0.34 nm, $\delta = 1.4$) and negative detuning (in the range from -0.15 nm ($\delta = -0.75$) to -0.10 nm ($\delta = -0.5$)). The respective two-state high- and low-state transmissions are illustrated in the inset of Figure 2b, for each switching cycle. With the negative detuning and the triangular pulses, the carrier-induced (Drude) blue-shifted dispersion overshoots the cavity resonance from the drive frequency and then thermally pins the cavity resonance to the laser drive frequency (detailed in Supplementary Information, Section S4). Since the free carrier lifetime of the hybrid media is about 200 ps and significantly lower than the drive pulse duration, these series of measurements are thermally dominated; the clear (attenuated) resonance dips on the intensity up-sweeps (down-sweeps) are due to the measurement sampling time shorter than the thermal relaxation timescale and a cooler (hotter) initial cavity temperature.

When the input laser intensity is well above the bistability threshold, the graphene-cavity system deviates from the two-state bistable switching and becomes oscillatory as shown in Figure 3a. Regenerative oscillation has only been suggested in a few prior studies, such as theoretically predicted in GaAs nanocavities with large Kerr nonlinearities [44] or observed in high-$Q$ ($3\times10^5$) silicon microdisks [45]. These regenerative oscillations are formed between the competing phonon and free carrier populations, with slow thermal red-shifts (~ 10 ns timescales) and fast free-carrier plasma dispersion blue-shifts (~ 200 ps timescales) in the case of our graphene-silicon cavities. The self-induced oscillations across the drive laser frequency are observed at threshold cavity powers of 0.4 mW, at ~ 9.4 ns periods in these series of measurements which gives ~ 106 MHz modulation rates, at experimentally-optimized detunings from $\delta_{(t=0)} = 0.68$ to 1.12. We emphasize that, for a monolithic silicon *L3* cavity, such regenerative pulsation has not been observed nor predicted to be observable at a relatively modest $Q$ of 7,500 (see Supplementary Information, Section S4), and attenuated by significant nonlinear absorption.

Figure 3b shows the input-output intensity cycles constructed from the temporal response measurements of a triangular-wave modulated 1.2 mW laser with a 2 µs cycle. Clear bistability behavior is seen below the carrier oscillation threshold. The system transits to the regime of self-sustained oscillations as the power coupled into the cavity is above the threshold, by tuning the laser wavelength into cavity resonance. We show an illustrative numerical modeling in Figure 3c: the fast free-carrier response fires the excitation pulse (blue dashed line; start cycle)), and heat diffusion (red solid line) with its slower time constant determines the recovery to the quiescent state in the graphene-cladded suspended silicon membrane. The beating rate between the thermal and free carrier population is around 50 MHz, as shown in the inset of Figure 3d, with the matched experimental data and coupled-mode theory simulation. The beating gives rise to tunable peaks in the radio frequency spectra (Figure 3d; blue solid line), which are absent when the input power is below the oscillation threshold (grey dashed line). As a supplementary detail, we note that the model does not include a time varying cavity quality factor, considering the high power would usually broaden the cavity bandwidth.

To examine only the Kerr nonlinearity, next we performed degenerate four-wave mixing measurements on the hybrid graphene – silicon photonic crystal cavities as illustrated in Figure 4, with continuous-wave laser input. A lower-bound $Q$ of 7,500 was specifically chosen to allow a ~ 200 pm cavity linewidth within which the highly dispersive four-wave mixing can be examined. The input pump and signal laser detunings are placed within this linewidth, with matched TE-like input polarization, and the powers set at 600 µW. Two example series of idler measurements are illustrated in Figure 4a and 4b, with differential pump and signal detunings respectively. In both series the parametric idler is clearly observed as a sideband to the cavity resonance, with the pump detuning ranging -100 pm to 30 pm and the signal detuning ranging from -275 pm to -40 pm, and from 70 pm to 120 pm (shown in Figure 4d). For each fixed signal- and pump-cavity detunings, the generated idler shows a slight intensity roll-off from linear signal (or pump) power dependence when the transmitted signal (or pump) power is greater than ~ 400 µW due to increasing free-carrier absorption effects (Supplementary Information, Figure S5). As illustrated in Figure 4a and 4b, the converted idler wave shows a four-wave mixing 3-dB bandwidth roughly matching the cavity linewidth when the pump laser is centered on the cavity resonance.

A theoretical four-wave mixing model with cavity field enhancement (Figure 4c and 4d) matches with these first graphene-cavity observations, and described in further detail in the Supplementary Information (Section S5). Based on the numerical model match to the experimental observations, the observed Kerr coefficient $n_2$ of the graphene-silicon cavity ensemble is $4.8 \times 10^{-17}$ m$^2$/W, an order of magnitude larger than in monolithic silicon and GaInP-related materials [24], and two orders of magnitude larger than in silicon nitride [25]. Independently we also modeled the field-averaged effective $\chi^{(3)}$ and $n_2$ of the hybrid graphene-silicon cavity, described as $\overline{n_2} = (\frac{\lambda_0}{2\pi})^d \frac{\int n^2(r) n_2(r) (|E(r) \cdot E(r)|^2 + 2|E(r) \cdot E(r)^*|^2) d^d r}{(\int n^2(r) |E(r)|^2 d^d r)^2}$,

where $E(r)$ is the complex fields in the cavity, $n(r)$ is local refractive index, $\lambda_0$ is the wavelength in vacuum, and $d$ is the number of dimensions (3). As detailed in the Supplementary Information (Section 5), the computed $n_2$ is at $7.7 \times 10^{-17}$ m$^2$/W, matching well with the observed four-wave mixing derived $n_2$. The remaining discrepancies arise from a Fermi velocity slightly smaller than the ideal values (~ $10^6$ m/s) in the graphene. As illustrated in Figure 4d for both measurement and theory, the derived conversion efficiencies are observed up to -30-dB in the unoptimized graphene-cavity, even at cavity $Q$s of 7,500 and low pump powers of 600 µW. The highly-doped graphene with Fermi-level level in the optical transparency region is a pre-requisite to these observations. We note that for a silicon cavity without graphene the conversion efficiencies are dramatically lower (by more than 20-dB lower) as shown in dashed black line, and even below the pump/signal laser spontaneous emission noise ratio (dotted grey line) preventing four-wave mixing observation in a single monolithic silicon photonic crystal cavity until now.

We have demonstrated for the first time a hybrid graphene – silicon optical cavity for chip-scale optoelectronics, with third-order nonlinear observations ranging from resonant optical bistability for optical signal processing at femtojoule level switching per bit, to temporal regenerative oscillations at record femtojoule cavity circulating powers for optically-driven and controlled reference oscillators, to graphene-cavity enhanced four-wave mixing at femtojoule energies on the chip. The transferred graphene on a wavelength-scale localized optical cavity enhances not only the thermal nonlinearities but also the ultrafast effective Kerr nonlinearity, suggesting a new parameter space for chip-scale optical physics and ultrafast optics in optical information processing.

**Methods**

**Device fabrication:** The photonic crystal nanostructures are defined by 248 nm deep-ultraviolet lithography in the silicon CMOS foundry onto undoped silicon-on-insulator (100) substrates. Optimized lithography and reactive ion etching was used to produce device lattice constants of 420 nm, hole radius of 124 ± 2 nm. The photonic crystal cavities and waveguides are designed and fabricated on a 250 nm silicon device thickness, followed by a buffered hydrofluoric wet-etch of the 1 μm buried oxide to achieve the suspended photonic crystal nanomembranes.

Centimeter-scale graphene are grown on 25 μm thick copper foils by chemical vapor deposition of carbon. The top oxide layer of copper is firstly removed in the hydrogen atmosphere (50 mTorr, 1000$^o$C, 2 sccm $H_2$ for 15 min), then monolayer carbon is formed on copper surface (250 mTorr, 1000$^o$C, 35 sccm $CH_4$, 2 sccm $H_2$ for 30 min). The growth is self-limiting once the carbon atom covers the Cu surface catalytic. The single layer graphene is then rapidly cooled down before being moved out of chamber. Poly-methyl-methacrylate (PMMA) is then spun-casted onto the graphene and then the copper foil etch-removed by floating the sample in $FeNO_3$ solution. After the metal is removed, graphene is transferred to a water bath before subsequent transfer onto the photonic crystal membranes. Acetone dissolves the PMMA layer, and the sample is rinsed with isopropyl alcohol and dry baked for the measurements.

**Optical measurements:** Continuous-wave finely-tuned semiconductor lasers from 1520 to 1620 nm (200 kHz bandwidth and -20 dBm to 7 dBm powers) were used for the measurements. Lensed tapered fibers (Ozoptics) with polarization controller and integrated on-chip spot size converters are used. Without the graphene cladding (in the control sample), the total fiber-chip-fiber transmission is ~ -10 dB. The fiber to channel waveguide coupling is optimized to be 3 dB per input/output facet, with 1 to 2 dB loss from channel to photonic crystal waveguide coupling. The linear propagation loss for our air-cladded photonic crystal waveguide has been determined at 0.6 dB/mm; for a photonic crystal waveguide length of 0.12 mm, the propagation loss in the waveguide is negligible. The output is monitored by an amplified InGaAs photodetector (Thorlabs PDA10CF, DC-150 MHz bandwidth) and oscilloscope (WaveJet 314A, 100 MHz bandwidth, 3.5 ns rise time) for the time-domain oscillations. The four-wave mixing pump laser linewidth is 10 pm (~ 1.2 GHz). Confocal microscopy was used for the graphene Raman spectroscopic measurements with a 100× (numerical aperture at 0.95) objective, pumped with a 514 nm laser.

**Numerical simulations:** Three-dimensional finite-difference-time-domain (FDTD) method with sub-pixel averaging is used to calculate the real and imaginary parts of the $E$-field distribution for the cavity resonant mode. The spatial resolution is set at 1/30 of the lattice constant (14 nm). Time-domain coupled mode theory, including free carrier dispersion and dynamics and thermal time constants, is carried out with 1 ps temporal resolution.

**Acknowledgments**

The authors acknowledge valuable discussions with Tony Heinz, along with Alexander Gondarenko, Felice Gesuele, Yilei Li, Joshua Lui, and Jinghui Yang for helpful suggestions. The authors acknowledge funding support from NSF IGERT (DGE-1069240) and the Center for Re-Defining Photovoltaic Efficiency through Molecule Scale Control, an Energy Frontier Research Center funded by the U.S. Department of Energy, Office of Science, Office of Basic Energy Sciences under Award Number DE-SC0001085.

T.G., J.F.M. performed the experiments, T.G., N.P., A.V.D.Z., J.H. prepared the graphene transfer and synthesis, M.Y. G.Q.L., D.L.K. nanofabricated the membrane samples, T.G., C.W.W. performed the numerical simulations, and T.G., C.W.W. prepared the manuscript.

**Additional information**

The authors declare no competing financial interests. Supplementary information accompanies this paper at www.nature.com/naturephotonics. Reprints and permission information is available online at http://www.nature.com/reprints/. Correspondence and requests for materials should be addressed to T.G. and C.W.W.

**Figure captions:**

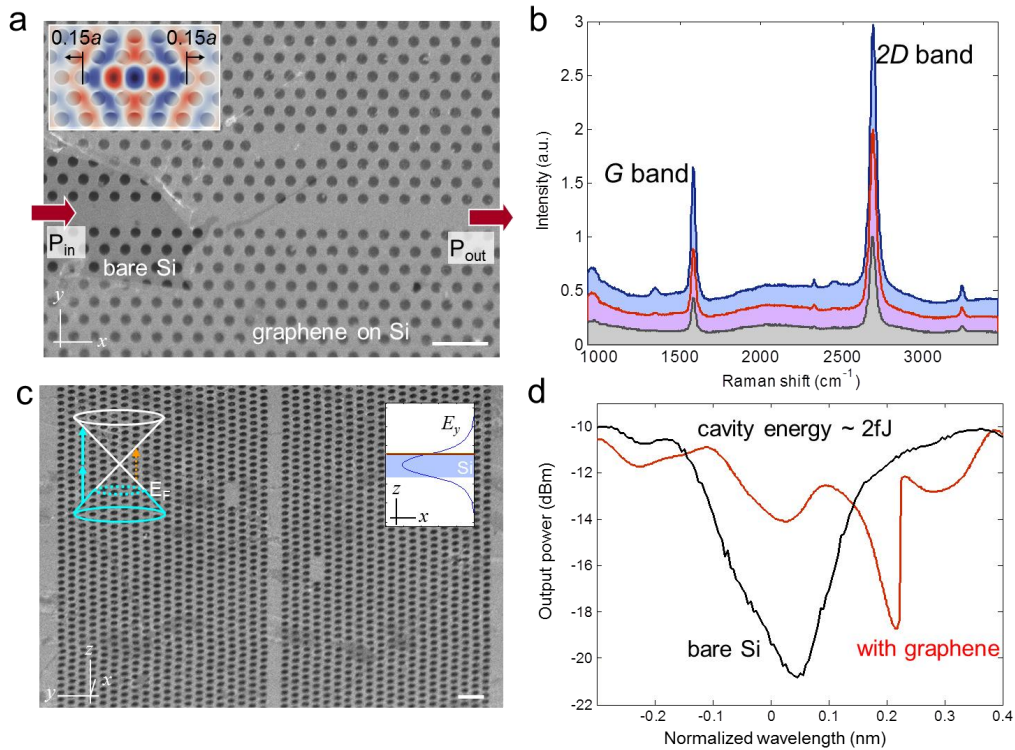

**Figure 1 | Graphene-cladded silicon photonic crystal nanostructures. a,** Scanning electron micrograph (SEM) of tuned photonic crystal cavity, with lattice constant *a* of 420 nm. Example SEM with separated graphene monolayer on silicon for illustration. Scale bar: 500 nm. Inset: example $E_z$-field from finite-difference time-domain computations. **b,** Measured Raman scattering spectra of monolayer CVD-grown graphene on photonic crystal cavity membrane. The Lorentzian lineshape full-width half-maximum of the *G* band (34.9 cm$^{-1}$) and *2D* band (49.6 cm$^{-1}$) peaks and the *G*-to-*2D* peak ratio indicates the graphene monolayer, while the single symmetric *G* peak indicates good graphene uniformity. Homogeneity across the sample is examined by excitation at different locations across the cavity membrane (blue, red and grey). **c,** SEM of suspended graphene-silicon membrane. Dark patches denote bilayer graphene. Left inset: Dirac

cone illustrating the highly-doped Fermi level (dashed blue circle) allowing only two-photon transition (blue arrows) while the one-photon transition (orange dashed arrow) is forbidden. Right inset: Computed $E_y$-field along $z$-direction, with graphene at the evanescent top interface. Scale bar: 500 nm. **d,** Example measured graphene-cladded cavity transmission with asymmetric Fano-like lineshapes (red dotted line) and significantly larger red-shift, compared to a control bare Si cavity sample with symmetric Lorentzian lineshapes (black dashed line). Both spectra are measured at 0.6 mW input power, and are centered to the intrinsic cavity resonances ($\lambda_{cavity\_0}$ = 1562.36 nm for graphene sample, and $\lambda_{cavity\_0}$ = 1557.72 nm for Si sample), measured at low power (less than 100 µW input power). The intrinsic cavity quality factor is similar between the graphene and the control samples.

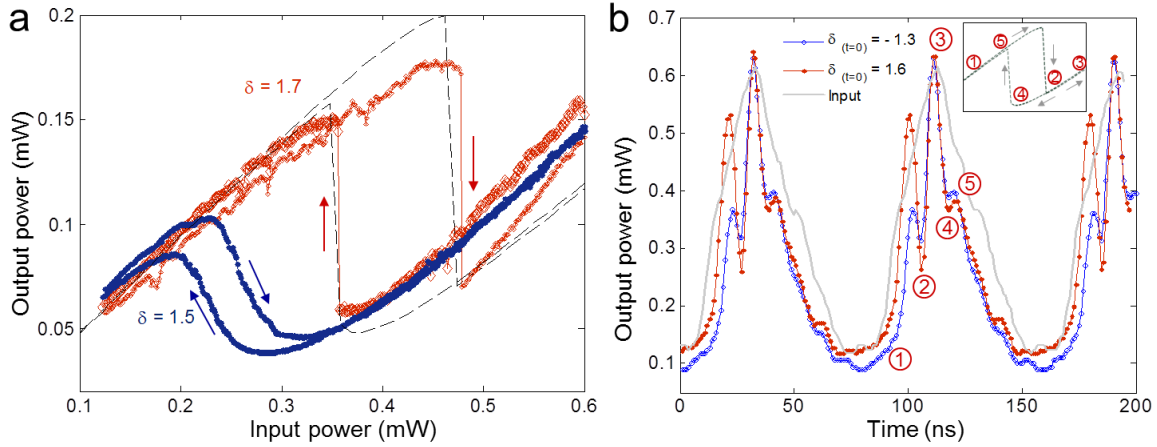

**Figure 2 | Bistable switching in graphene-cladded nanocavities. a**, Steady-state input/output optical bistability for the quasi-TE cavity mode with laser-cavity detuning $\delta$ at 1.5 ($\lambda_{laser}$ = 1562.66 nm) and 1.7 ($\lambda_{laser}$ = 1562.70 nm). The dashed black line is the coupled-mode theory simulation with effective nonlinear parameters of the graphene-silicon cavity sample. **b,** Switching dynamics with triangular waveform drive input (dashed grey line). The bistable resonances are observed for both positive and negative detuning. Blue empty circles: $\delta(\text{t=0})$ = -1.3($\lambda_{laser}$ = 1562.10 nm), red solid circles: $\delta(\text{t=0})$ = 1.6 ($\lambda_{laser}$ = 1562.68 nm). Inset: schematic of high- and low-state transmissions.

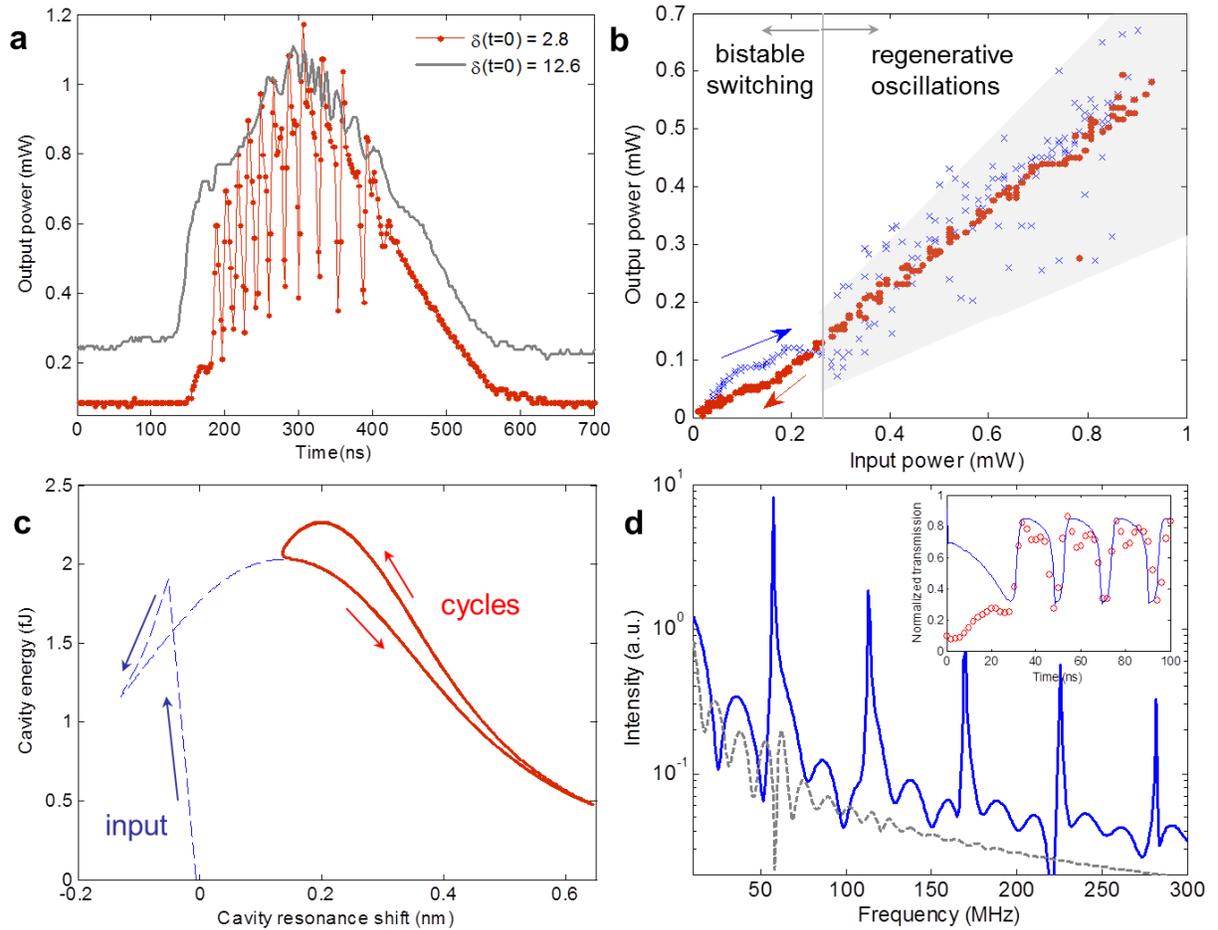

**Figure 3 | Regenerative oscillations in graphene-cladded nanocavities. a**, Observations of temporal regenerative oscillations in the cavity for optimized detuning ($\lambda_{laser}$ = 1562.47 nm). The input power is quasi-triangular waveform with peak power 1.2 mW. The grey line is the reference output power, with the laser further detuned at 1.2 nm from cavity resonance ($\lambda_{laser}$ =1563.56 nm). **b,** Mapping the output power versus input power with slow up (blue cross) and down (red) power sweeps. In the up-sweep process, the cavity starts to oscillate when the input power is beyond 0.29 mW. **c,** Nonlinear coupled-mode theory model of cavity transmission versus resonance shift, in the regime of regenerative oscillations. With a detuning of 0.15 nm [$\delta(t=0)$ = 0.78] the free carrier density swings from 4.4 to 9.1 ×$10^{17}$ per $cm^3$ and the increased

temperature $\Delta T$ circulates between 6.6 and 9.1K. **d,** RF spectrum of output power at below (0.4 mW, grey dashed line) and above oscillation threshold (0.6 mW, blue solid line) at the same detuning $\delta_{(t=0)} = 0.78$ ($\lambda_{laser}$-$\lambda_{caviy}$ = 0.15 nm), as in panel **c**. Inset: Normalized transmission from model (blue line) and experimental data at the same constant power level (red circles).

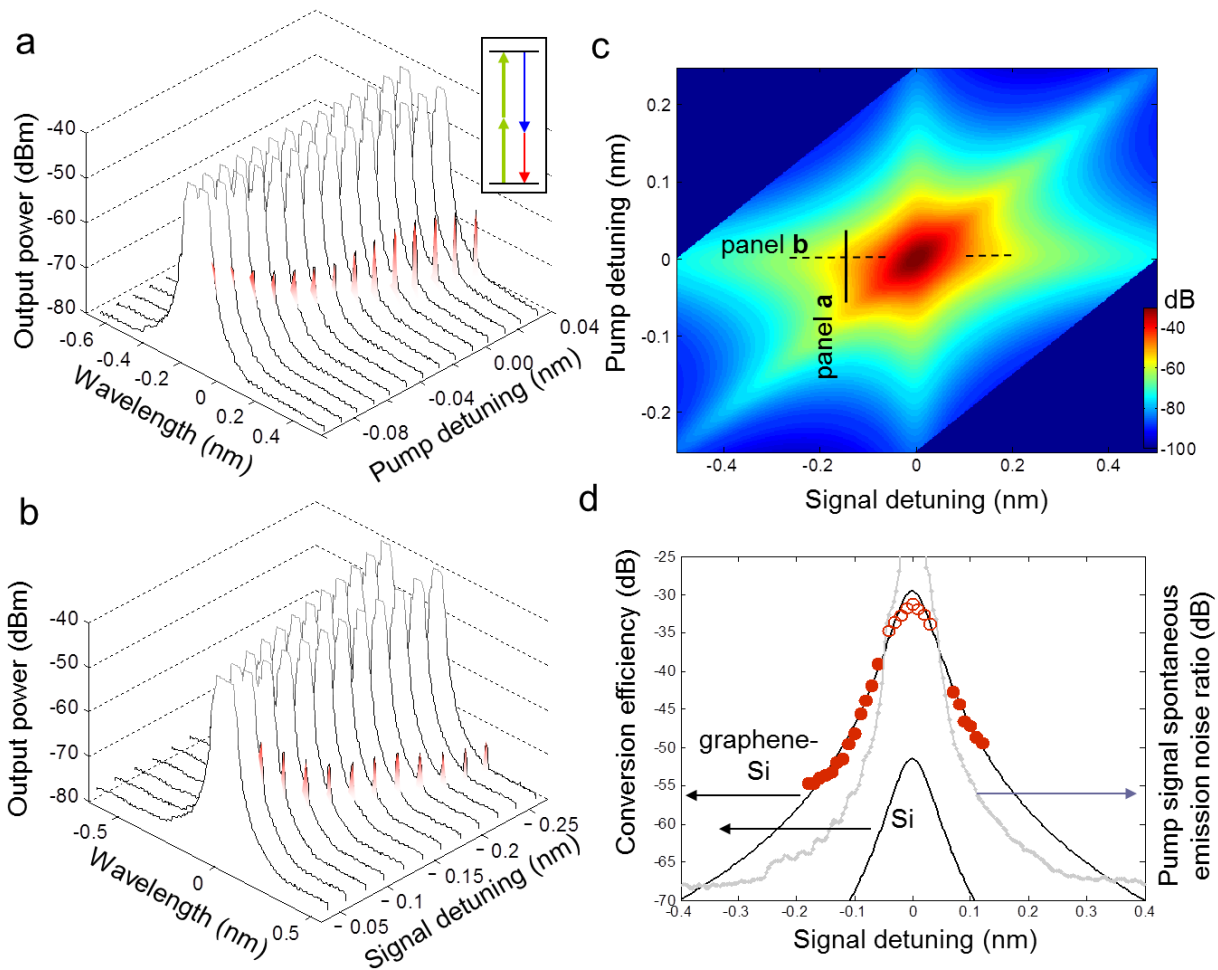

**Figure 4 | Parametric four-wave mixing in graphene-cladded silicon nanocavities. a**, Measured transmission spectrum with signal laser fixed at -0.16 nm according to cavity resonance, and pump laser detuning is scanned from -0.1 to 0.03 nm. Inset: band diagram of degenerate four-wave mixing process with pump (green), signal (blue) and idler (red) lasers. **b,** Measured transmission spectrum with pump laser fixed on cavity resonance, and signal laser detuning is scanned from -0.04 to -0.27 nm. **c,** Modeled conversion efficiency versus pump and signal detuning from the cavity resonance. The solid and dashed lines mark the region plotted in panel **a** and **b** respectively. **d,** Observed and simulated conversion efficiencies of the cavity. Red

solid dots are measured with signal detuning as in panel **b**, and the empty circles are obtained through pump detuning as in panel **a**, plus 29.5-dB (offset due to the 0.16 nm signal detuning). Solid and dashed black lines are modeled conversion efficiencies of graphene-silicon and monolithic silicon cavities respectively. Grey dashed line (superimposed): illustrative pump/signal laser spontaneous emission noise ratio.

Supplementary Information

S1. Dynamic conductivity and optical absorption of graphene

**S1.A. Estimating the Fermi level in CVD grown graphene**

The Raman spectra are shown in Figure 1b and Figure S1a. The G and 2D band peaks are excited by a 514 nm laser and are located at 1582 cm$^{-1}$ and 2698 cm$^{-1}$ respectively. The Raman spectra are homogeneous within one device, and vary less than 5 cm$^{-1}$ from sample to sample. The Lorentzian lineshape with full-width half-maximum of the G (34.9 cm$^{-1}$) and 2D (49.6 cm$^{-1}$) band indicates the graphene monolayer [S1], perhaps broadened by chemical doping and disorder. The phonon transport properties are represented by the G and 2D peak positions (varying within 1 cm$^{-1}$ over the sample) and the intensity ratios between the G and 2D peaks (fluctuating from 1 to 1.5) [S2].

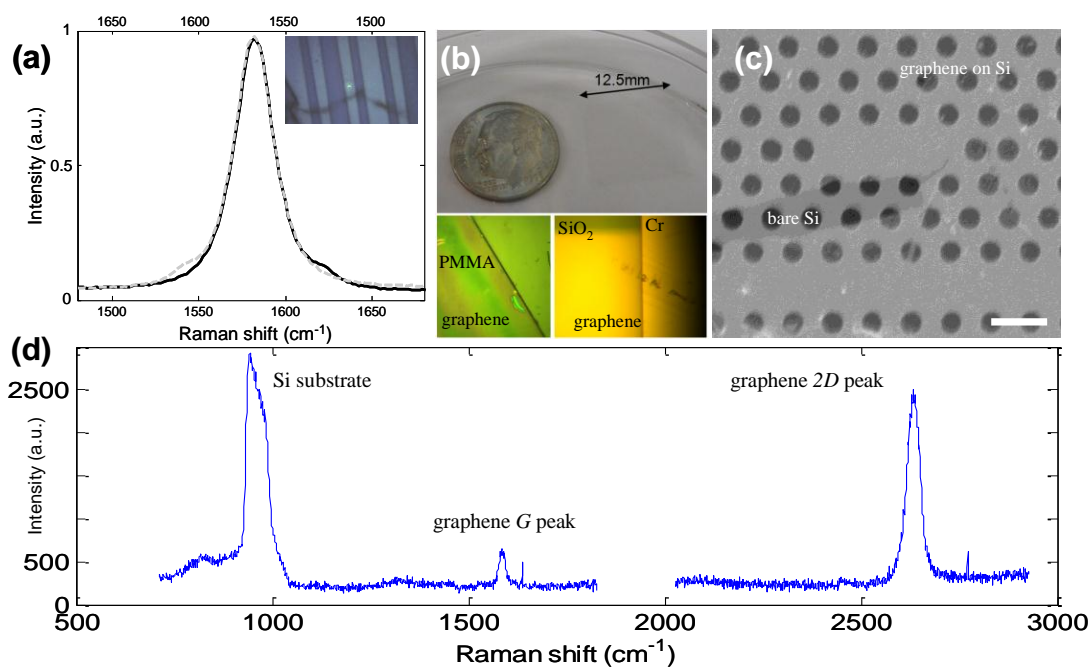

**Figure S1 | Raman spectrum and transferred graphene samples. a,** Raman G peak (black line) and its inverse (grey dashed line) to illustrate G peak symmetry. Inset: optical micrograph of the device with graphene transferred under Raman measurement. **b,** A centimeter-scale graphene film prepared. Optical micrograph of graphene film transferred to various substrates

(PMMA), air-bridged silicon membranes, silicon oxide and partially covered metal surfaces, with graphene interface pictured. **c,** Scanning electronic micrograph of example air-bridged device sample with graphene covering the whole area except the dark (exposed) region. Scale bar: 500 nm. **d,** Complete Raman spectrum of the graphene-clad silicon membrane samples.

Figure S1b and S1c illustrates example transfers of large-area CVD graphene into various substrates including air-bridged silicon membranes, silicon oxide, and partially covered metal surfaces. CVD grown graphene is thicker and has rough surfaces compared to exfoliated graphene, shown by the broadened 2D peak and the fluctuation of the 2D versus G peak ratio [S3]. The thickness of graphene is ~ 1 nm. Wrinkles on the graphene surface are formed during the cool down process, due to the differential thermal expansion between the copper substrate and graphene, and consistently appear only at the edges of our samples. We emphasize that at the device regions most of the devices are covered with a single unwrinkled graphene layer.

The 2D peak is observable only when the laser excitation energy ($E_L$) and the energy corresponding to electron-hole recombination process ($E_T$) follow the relation: $(E_L-E_T)/2 > E_F$, where $E_F$ is the Fermi energy of graphene. With 514 nm laser excitation, the *2D* peak is located at 2698 cm$^{-1}$ (Figure 1b and Figure S1a).

We note that wet transfer of graphene is used in these measurements. While a very thin (in the range of nanometers) residual layer of PMMA can remain on the sample after transfer, PMMA typically only has a non-centrosymmetric $\chi^{(2)}$ response with a negligible $\chi^{(3)}$ response and hence does not contribute to the enhanced four-wave mixing observations. The dopants can arise from residual absorbed molecules or ions on graphene or at the grain boundaries, during the water bath and transfer process. With the same CVD growth process, we also examined the dry transfer technique which controls the doping density to be low enough such that the Fermi level is within the interband optical transition region. In that case, the measured samples have a significantly increased propagation loss from ~0dB to ~ 11 dB over the 120 μm length photonic crystal waveguide. The wet transfer technique significantly reduced the linear absorption, thereby allowing the various nonlinear optoelectronic measurements observed in this work.

## S1.B. Calculations of graphene's dynamic conductivity

Given the fact that CVD graphene is heavily p-doped, the dynamic conductivity for intra- and inter-band optical transitions [S4] can be determined from the Kubo formalism as:

$$\sigma_{intra}(\omega) = \frac{je^2\mu}{\pi\hbar(\omega+j\tau^{-1})} \quad , \tag{S-E1}$$

$$\sigma_{inter}(\omega) = \frac{je^2\mu}{4\pi\hbar}\ln(\frac{2|\mu|-\hbar(\omega+j\tau^{-1})}{2|\mu|+\hbar(\omega+j\tau^{-1})}), \tag{S-E2}$$

where $e$ is the electron charge, $\hbar$ is the reduced Plank constant, $\omega$ is the radian frequency, $\mu$ is chemical potential, and $\tau$ is the relaxation time (1.2 ps for interband conductivity and 10 fs for intraband conductivity). The dynamic conductivity of intra- and inter-band transitions at 1560 nm are $(-0.07-0.90i)\times10^{-5}$ and $(4.15-0.95i)\times10^{-5}$ respectively, leading to the total dynamic conductivity $\sigma_{total}=\sigma_{intra}+\sigma_{inter}$ of $(4.1-1.8i)\times10^{-5}$. Given negative imaginary part of total conductivity, the TE mode is supported in graphene [S5]. The light can travel along the graphene sheet with weak damping and thus no significant loss is observed for the quasi-TE mode confined in the cavity [S6]. The impurity density of the 250 nm silicon membrane is $\sim10^{11}$ cm$^{-2}$, slightly lower than the estimated doping density in graphene.

## S2. Parameter space of nonlinear optics in graphene nanophotonics

Figure S2 compares cavity-based switching and modulation across different platforms including silicon, III-V and the hybrid graphene- silicon cavities examined in this work. The thermal or free-carrier plasma-based switching energy is given by $P_{0th/e}\times\tau_{th/e}$, where $P_{0th/e}$ is the threshold laser power required to shift the cavity resonance half-width through thermal or free-carrier dispersion; $\tau_{th/e}$ are the thermal and free-carrier lifetimes in resonator. Note that the lifetime should be replaced by cavity photon lifetime if the latter is larger (for high $Q$ cavity). Graphene brings about a lower switching energy due to strong two-photon absorption (~3,000 cm/GW) [S7]. The recovery times of thermal switching (in red) are also shortened due to higher thermal conductivity in graphene, which is measured for supported graphene monolayers at 600 W/mK [S8] and bounded only by the graphene-contact interface and strong interface phonon scattering.

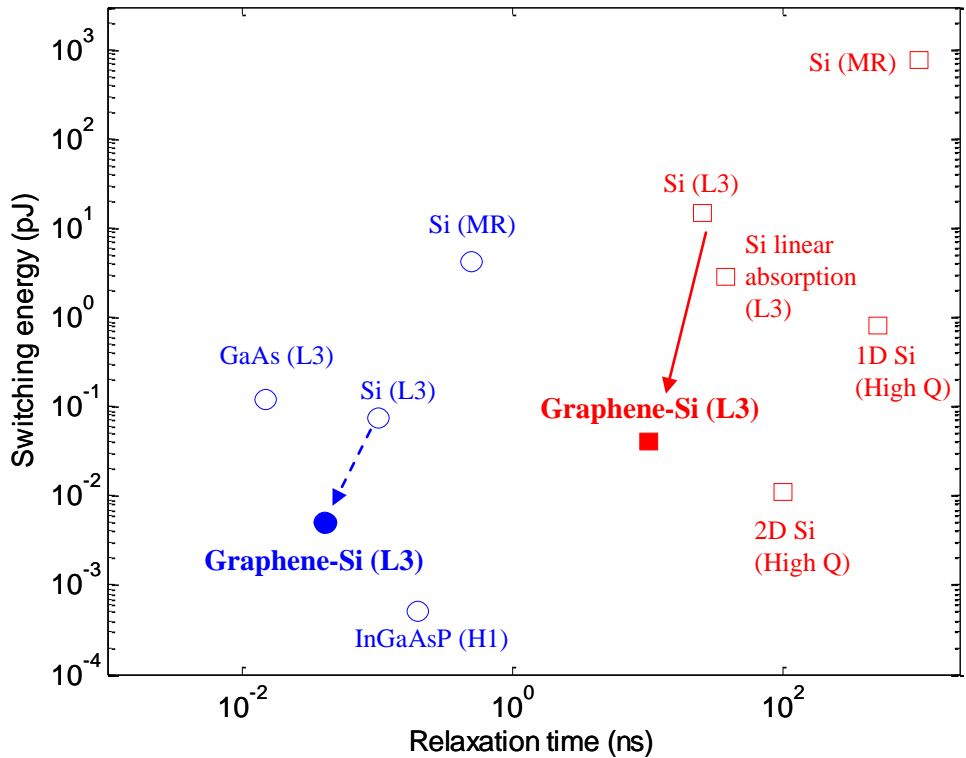

**Figure S2 | Comparison of switching energy versus recovery time of cavity-based modulators and switches across different semiconductor material platforms.** The blue circles are carrier plasma-induced switches with negative detuning, and the red squares are thermal-optic switches with positive detuning. The dashed lines illustrate the operating switch energies versus recovery times, for the same material [S9-16]. *L3* (*H1*) denotes photonic crystal *L3* (*H1*) cavity; *MR* denotes microring resonator.

The switching energy is inversely proportional to two photon absorption rate ($β_2$). Table I below summarizes the first-order estimated physical parameters from: (1) coupled-mode theory and experimental data matching; (2) full three-dimensional numerical field simulations, and (3) directly measured data, further detailed in the various sections of this Supplementary Information. With the enhanced two-photon absorption in graphene and first-order estimates of the reduced carrier lifetimes (detailed in Section S3), the switching energy – recovery time performance of the hybrid graphene-silicon cavity is illustrated in Figure S2, compared to monolithic GaAs or silicon ones.

TABLE I | **Estimated physical parameters from time-dependent coupled-mode theory-experimental matching, three-dimensional numerical field simulations, and measurement data.**

| Parameter | Symbol | GaAs [S17] | Si | Monolayer Graphene-Si |
|---|---|---|---|---|
| TPA coefficient | $\beta_2$ (cm/GW) | 10.2 | 1.5 [S18] | 25 [3D] |
| Kerr coefficient | $n_2$ (m$^2$/W) | $1.6 \times 10^{-17}$ | $0.44 \times 10^{-17}$ [S18] | $7.7 \times 10^{-17}$ [3D] |
| Thermo-optic coeff. | $dn/dT$ | $2.48 \times 10^{-4}$ | $1.86 \times 10^{-4}$ | |
| Specific heat | $c_v \rho$ (W/Km$^{-3}$) | $1.84 \times 10^6$ | $1.63 \times 10^6$ [cal] | |
| Thermal relaxation time | $\tau_{th,c}$ (ns) | 8.4 | 12 | 10 [cal] |
| Thermal resistance | $R_{th}$ (K/mW) | 75 | 25 [19] | 20 [cal] |
| FCA cross section | $\sigma$ ($10^{-22}$ m$^3$) | 51.8 | 14.5 | |
| FCD parameter | $\zeta$ ($10^{-28}$ m$^3$) | 50 | 13.4 | |
| Carrier lifetime | $\tau_{fc}$ (ps) | 8 | 500 [S20] | 200 [CMT] |
| Loaded Q | Q | 7000 | 7000 [m] | |
| Intrinsic Q | $Q_0$ | 30,000 | 23,000 [m] | |

[CMT]: nonlinear time-dependent coupled-mode theory simulation; [3D]: three-dimensional numerical field calculation averages; [m]: measurement at low power; [cal]: first-order hybrid graphene-silicon media calculations. $\tau_{fc}$ is the effective free-carrier lifetime accounting for both recombination and diffusion.

S3. Graphene two-photon absorption and accompanying thermal and free-carrier nonlinearities

With increasing input power, the transmission spectra evolve from symmetric Lorentzian to asymmetric lineshapes as illustrated in the examples of Figure 1d and Figure S3. Through second-order perturbation theory [S7], the two-photon absorption coefficient $\beta_2$ in monolayer graphene is estimated through the second-order interband transition probability rate per unit area as:

$$\beta_2 = \frac{4\pi^2}{\varepsilon_\omega \omega^4 \hbar^3} \left( \frac{v_F e^2}{c} \right)^2 , \qquad \text{(S-E3)}$$

where $v_F$ is the Fermi velocity, $\hbar$ is the reduced Planck's constant, $e$ is the electron charge, and $\varepsilon_\omega$ is the permittivity of graphene in the given frequency. At our 1550 nm wavelengths, $\beta_2$ is determined through Z-scan measurements and first-principle calculations to be in the range of ~ 3,000 cm/GW [S7].

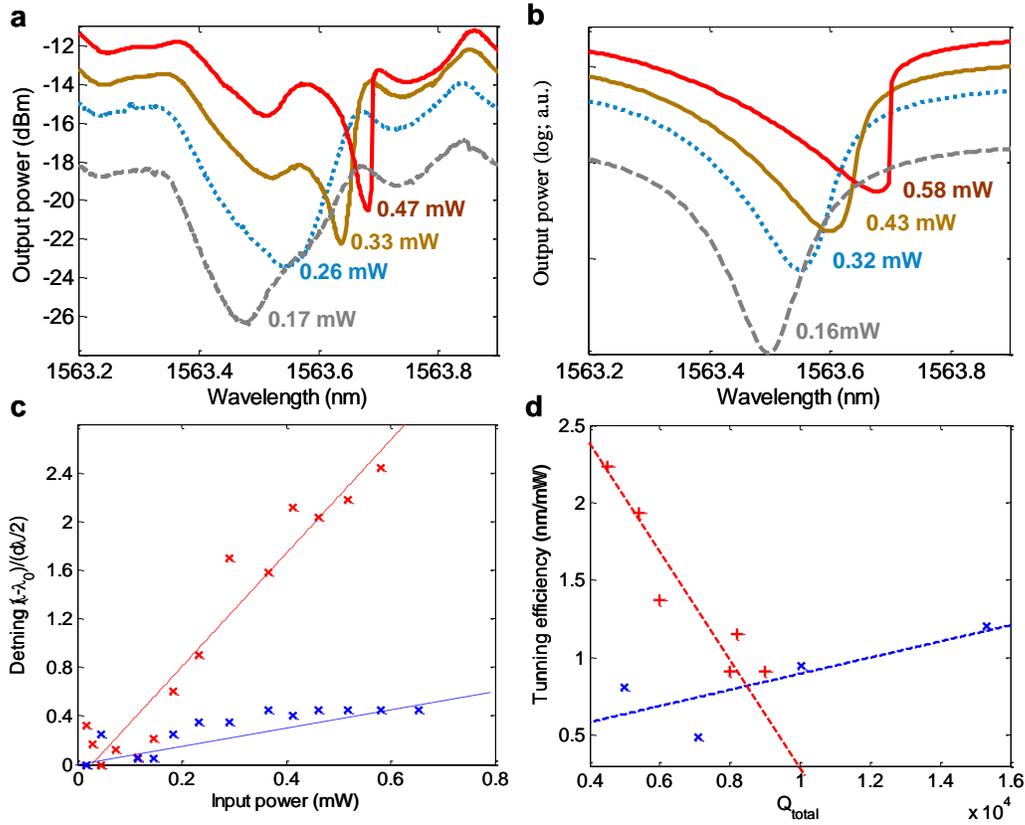

**Figure S3 | Steady-state two-photon absorption induced thermal nonlinearities in graphene-silicon hybrid cavities. a,** Measured quasi-TE transmission spectra of a graphene-clad *L3* cavity with different input power levels (with extracted insertion loss from the facet of waveguides in order to be comparable to simulation in b). **b,** Nonlinear coupled-mode theory simulated transmission spectra. The estimated input powers are marked in the panels. **c,** Measured cavity resonance shifts versus input power, with the graphene-clad cavity samples (in red) and the monolithic silicon control cavity sample (in blue). **d,** Tuning efficiencies for graphene-clad cavity samples (in red) and control cavity samples (in blue) for a range of cavity loaded *Q*-factors examined.

We model the nonlinear cavity transmissions with time-domain nonlinear coupled-mode theory for the temporal rate evolution of the photon, carrier density and temperatures as described by [S21]:

$$\frac{da}{dt} = (i(\omega_L - \omega_0 + \Delta\omega) - \frac{1}{2\tau_t})a + \kappa\sqrt{P_{in}}, \quad \text{(S-E4)}$$

$$\frac{dN}{dt} = \frac{1}{2\hbar\omega_0\tau_{TPA}}\frac{V_{TPA}}{V_{FCA}^2}|a|^4 - \frac{N}{\tau_{fc}}, \quad \text{(S-E5)}$$

$$\frac{d\Delta T}{dt} = \frac{R_{th}}{\tau_{th}\tau_{FCA}}|a|^2 + \frac{\Delta T}{\tau_{th}}, \quad \text{(S-E6)}$$

where $a$ is the amplitude of resonance mode; $N$ is the free-carrier density; $\Delta T$ is the cavity temperature shift. $P_{in}$ is the power carried by incident continuous-wave laser. $\kappa$ is the coupling coefficient between waveguide and cavity, adjusted by the background Fabry-Perot resonance in waveguide [S22]. $\omega_L$-$\omega_0$ is the detuning between the laser frequency ($\omega_L$) and cold cavity resonance ($\omega_0$). The time-dependent cavity resonance shift is $\Delta\omega=\Delta\omega_N-\Delta\omega_T+\Delta\omega_K$, where the free-carrier dispersion is $\Delta\omega_N=\omega_0\zeta N/n$. The thermal induced dispersion is $\Delta\omega_T=\omega_0\Delta T(dn/dT)/n$. $\Delta\omega_K$ is the Kerr dispersion, and is negligibly small compared to the thermal and free-carrier mechanisms.

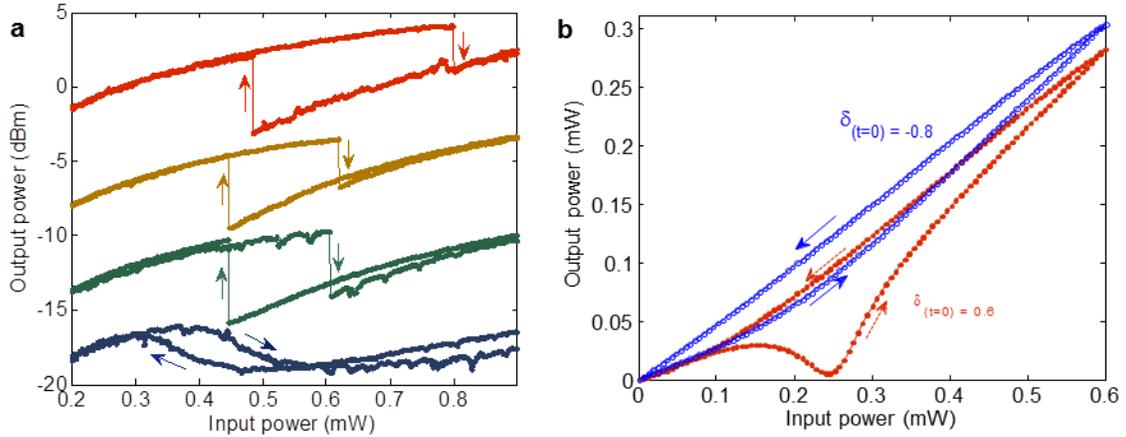

**Figure S4 | Bistable switching in graphene-clad nanocavities. a,** Measured steady-state bistability at different detunings set at 0.18, 0.23, 0.26, 0.29 nm (from bottom to top). The plots are offset for clarity: green (offset 2 dB), brown (offset 8 dB) and red lines (offset 15 dB). **b,** Coupled-mode equations calculated switching dynamics with triangular input. The output power versus input power for the positive (red) and negative (blue) detuning with triangular input.

The total loss rate is $1/\tau_t = 1/\tau_{in}+1/\tau_v+1/\tau_{lin}+1/\tau_{TPA}+1/\tau_{FCA}$. $1/\tau_{in}$ and $1/\tau_v$ is the loss rates into waveguide and vertical radiation into the continuum, ($1/\tau_{in/v}=\omega/Q_{in/v}$), the linear absorption $1/\tau_{lin}$ for silicon and graphene are demonstrated to be small. The free-carrier absorption rate $1/\tau_{FCA}=c\sigma N(t)/n$. The field averaged two-photon absorption rate $1/\tau_{TPA}=\overline{\beta_2}\,c^2/n^2/V_{TPA}/|a|^2$, where the effective two-photon absorption coefficient is defined as:

$$\overline{\beta_2} = (\frac{\lambda_0}{2\pi})^d \frac{\int n^2(r)\beta_2(r)(|E(r)\cdot E(r)|^2 + 2|E(r)\cdot E(r)^*|^2)d^d r}{(\int n^2(r)|E(r)|^2 d^d r)^2}, \quad \text{(S-E7)}$$

The mode volume for two-photon absorption (same as Kerr):

$$V_{TPA/Kerr} = \frac{(\int n^2(r)|A(r)|^2 dr^3)^2}{\int_{Si} n^4(r)|A(r)|^4 dr^3}, \quad \text{(S-E8)}$$

The effective mode volume for free-carrier absorption is:

$$V^2_{FCA} = \frac{(\int n^2(r)|A(r)|^2 dr^3)^3}{\int_{Si} n^6(r)|A(r)|^6 dr^3}. \quad \text{(S-E9)}$$

The model shows remarkable match to the measured transmissions. With the two-photon absorption and Kerr (Supplementary Information, Section 5) coefficients of the hybrid cavity calculated from 3D finite-difference time-domain field averages and first-order estimates of the thermal properties (specific heat, effective thermal resistance, and relaxation times), the carrier lifetime of the graphene-clad photonic crystal cavity is estimated to first-order at 200 ps.

## S4. Switching dynamics and regenerative oscillation in graphene-clad silicon cavities

From the nonlinear coupled-mode modeling, the dynamical responses of the hybrid cavity to step inputs are shown in Figure S5a, illustrating the switching dynamics and regenerative oscillations. Free-carrier dispersion causes the switching on the negative- detuned laser, and the thermal nonlinearity leads to the switching on the positive side. The interplay of the free-carrier-induced cavity resonance blue-shift dynamics with the thermal-induced cavity red-shift time constants is observed. Figure S5b shows the correspondent radio frequency spectrum. By tuning the laser wavelength, the fundamental mode can be set from 48 MHz (zero detuning) to 55 MHz (0.3 nm detuning). The dependence of oscillation period to the detuning and input laser power is further provided in Figure S5c and Figure S5d respectively.

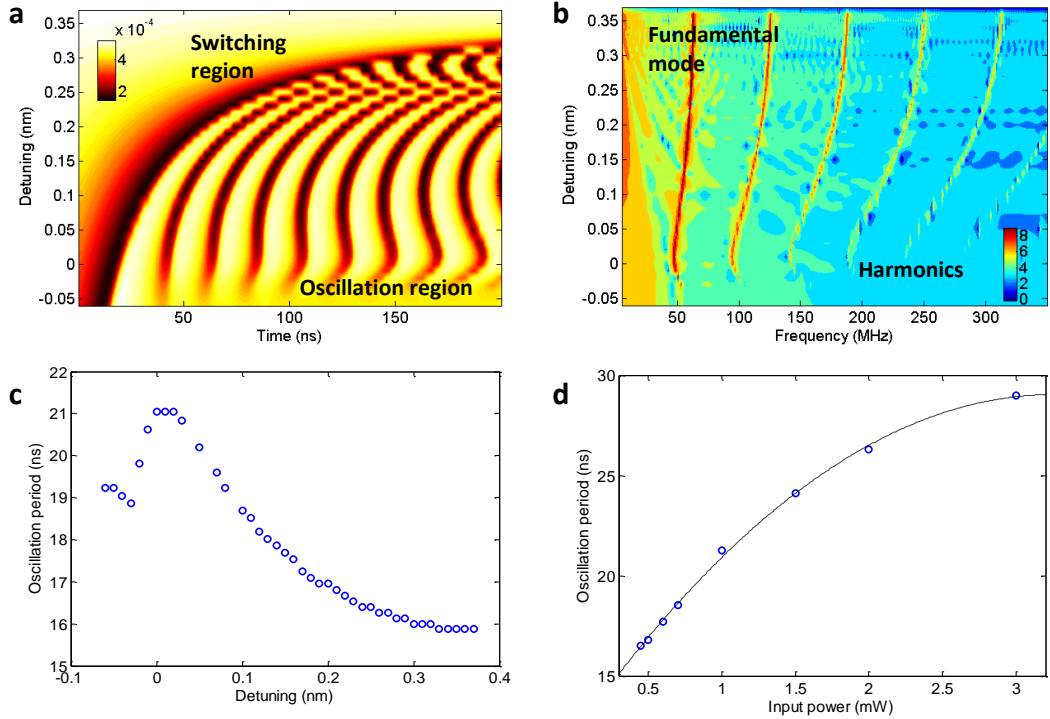

**Figure S5 | Coupled-mode equations calculated time domain response to a step input with a graphene-clad silicon photonic crystal *L3* nanocavity side-coupled to a photonic crystal waveguide. a,** The output versus input powers for positive and negative detunings (laser-cavity detunings are set from -0.06 to 0.37 nm). Input laser power is set at 0.6 mW. The cavity switching dip is observed for all detunings, and regenerative oscillation exists only predominantly for positive detuning. **b,** Frequency response of the cavity switching and oscillation dynamics with conditions as in **a** (in log scale). The laser detuning is set from -0.06 to 0.37 nm. **c** and **d.** Oscillation period versus laser detunings and input powers respectively.

Regenerative oscillations were theoretically predicted in GaAs nanocavities with large Kerr nonlinearities [S23], or observed only in high-*Q* silicon microdisks (*Q* at $3\times10^5$) with *V* at $40(\lambda/n_{Si})^3$, at sub-mW power levels [S24]. The graphene-enhanced two-photon absorption, free-carrier and thermal effects allow regenerative oscillations to be experimentally observable with $Q^2/V$ values [of $4.3\times10^7(\lambda/n)^3$] at least 50× lower, at the same power threshold levels. The regenerative oscillations with lower *Q*s allow higher speed and wider bandwidth operation, and are less stringent on the device nanofabrication.

## S5. Ultrafast Kerr in graphene – silicon hybrid structures

### S5.A. Computations of effective Kerr nonlinearity in graphene-clad silicon cavities

Third-order nonlinearity susceptibility for graphene is reported as large as $|\chi^{(3)}| \sim 10^{-7}$ esu in the wavelength range of 760 to 840 nm [S25]. When two external beams with frequency $\omega_1$ (pump) and $\omega_2$ (signal) are incident on graphene, the amplitude of sheet current generated at the harmonics frequencies ($2\omega_1-\omega_2$) is described by:

$$j_e = -\frac{3}{32}\frac{e^2}{\hbar}\varepsilon_2 \left(\frac{ev_F\varepsilon_1}{\hbar\omega_1\omega_2}\right)^2 \frac{2\omega_1^2 + 2\omega_1\omega_2 - \omega_2^2}{\omega_1(2\omega_1-\omega_2)}, \qquad \text{(S-E10)}$$

where $\varepsilon_1$, $\varepsilon_2$ are the electric field amplitudes of the incident light at frequencies $\omega_1$ and $\omega_2$ respectively. $v_F$ ($=10^6$ m/s) is the Fermi velocity of graphene. Under the condition that both $\omega_1$ and $\omega_2$ are close to $\omega$, the sheet conductivity can be approximated as:

$$\sigma^{(3)} = \frac{j_e}{\varepsilon_1\varepsilon_1\varepsilon_2} = -\frac{9}{32}\frac{e^2}{\hbar}\left(\frac{ev_F}{\hbar\omega^2}\right)^2, \qquad \text{(S-E11)}$$

Since most of the sheet current is generated in graphene, the effective nonlinear susceptibility of the whole membrane can be expressed as:

$$\chi^{(3)} = \frac{\sigma^{(3)}}{\omega d} = -\frac{9}{32}\frac{e^4 v_F^2}{\hbar^3 c^5}\frac{\lambda^5}{d}, \qquad \text{(S-E12)}$$

where d is the thickness of the graphene (~1 nm), $\lambda$ is the wavelength, and c is the speed of light in vacuum. The calculated $\chi^{(3)}$ of a monolayer graphene is in the order of $10^{-7}$ esu (corresponding to a Kerr coefficient $n_2 \sim 10^{-13}$ m$^2$/W), at $10^5$ times higher than in silicon ($\chi^{(3)} \sim 10^{-13}$ esu, $n_2 \sim 4\times10^{-18}$ m$^2$/W) [S26].

Effective $n_2$ of the hybrid graphene-silicon membrane is then calculated for an inhomogeneous cross-section weighted with respect to field distribution [S27]. With a baseline model without complex graphene-surface electronic interactions, the effective $n_2$ can be expressed as:

$$\overline{n_2} = \left(\frac{\lambda_0}{2\pi}\right)^d \frac{\int n^2(r)n_2(r)(|E(r)\cdot E(r)|^2 + 2|E(r)\cdot E(r)^*|^2)d^d r}{\left(\int n^2(r)|E(r)|^2 d^d r\right)^2}, \qquad \text{(S-E13)}$$

where $E(r)$ is the complex fields in the cavity and n(r) is local refractive index. The local Kerr coefficient $n_2(r)$ is $3.8 \times 10^{-18}$ m$^2$/W in silicon membrane and $\sim 10^{-13}$ m$^2$/W for graphene, $\lambda_0$ is the wavelength in vacuum, and d=3 is the number of dimensions. The complex electric field E(r) is obtained from 3D finite-difference time-domain computations of the optical cavity examined [S28]. The resulting field-balanced effective $n_2$ is calculated to be $7.7 \times 10^{-17}$ m$^2$/W ($\chi^{(3)} \sim 10^{-12}$esu), close to the best reported chalcogenide photonic crystal waveguides [S29, S30].

**TABLE II | Field-balanced third-order nonlinear parameter.**

| Computed parameters | $\overline{n_2}$ (m$^2$/W) | $\overline{\beta_2}$ (m/W) |
|---|---|---|
| Graphene | $10^{-13}$ [S25] | $10^{-7}$ [S18] |
| Silicon | $3.8 \times 10^{-18}$ | $8.0 \times 10^{-12}$ |
| Monolayer graphene-silicon | $7.7 \times 10^{-17}$ | $2.5 \times 10^{-11}$ |
| Chalcogenide waveguide | $7.0 \times 10^{-17}$ | $4.1 \times 10^{-12}$ |

Likewise, the effective two-photon absorption coefficient is computed in the same field-balanced approach, with a result of $2.5 \times 10^{-11}$ m/W. The resulting nonlinear parameter $\gamma$ (=$\omega n_2/cA_{eff}$) is derived to be 800 W$^{-1}$m$^{-1}$, for an effective mode area of 0.25 μm$^2$.

**S5.B. Local four-wave mixing in graphene-clad photonic crystals cavities**

The conversion efficiency of the single cavity $\eta = |\gamma P_p L'|^2 FE_p^4 FE_s^2 FE_c^2$, where $FE_p$, $FE_s$, and $FE_c$ are the field enhancement factors of pump, signal and idler respectively [S31]. The effective length L' includes the phase mismatch and loss effects. Compared to the original cavity length (~ 1582.6 nm), the effective cavity length is only slightly modified by less than 1 nm. However, the spectral dependent field enhancement factor is the square of the cavity build-up factor $FE^2 = P_{cav}/P_{wg} = F_{cav}(U/U_{max})\eta_p^2$, where U/U$_{max}$ is the normalized energy distribution with the Lorenzian lineshape. $\eta_p$=0.33 is the correction term for the spatial misalignment between the quasi-TE mode and graphene, and the polarization. The field enhancement effect in the cavity is proportional to the photon mode density: $F_{cav} = Q\lambda^3/(8\pi V)$ [S32].

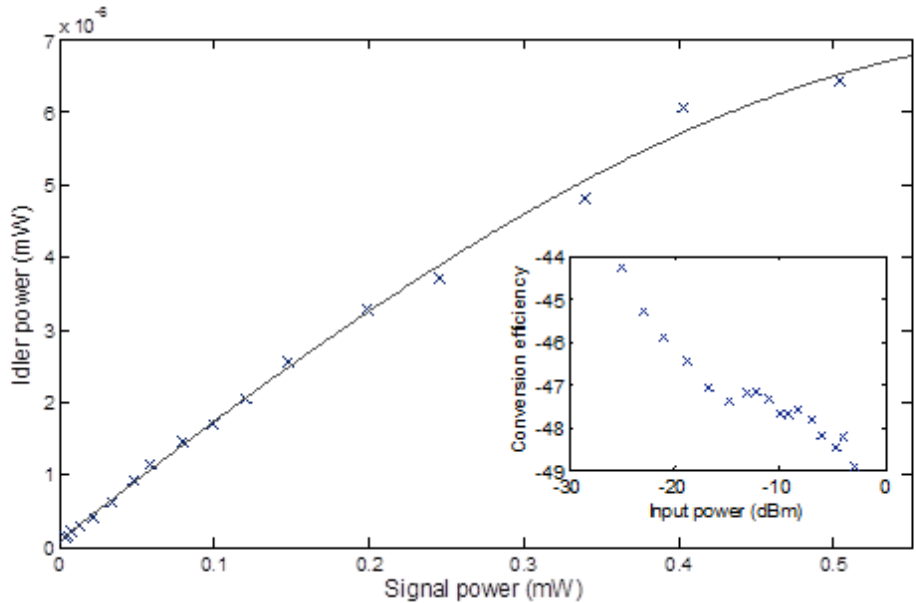

**Figure S6 | Free-carrier absorption effects on the four-wave mixing conversion efficiency.** Measured idler power versus signal power at the transmitted port, with the pump power is fixed on the cavity resonance and the the signal laser detuned by 200 pm. Experimental data (×) and quadratic fit (solid line). Inset: corresponding conversion efficiency versus signal power.

The enhanced two-photon-absorption and induced free-carrier absorption would produce nonlinear loss. To investigate the direct effect of two-photon absorption and free-carrier absorption on the four wave mixing, we measure the conversion efficiency with varying input signal power as shown in Figure S6. Extra 4-dB loss is measured when the input signal power increases from -22 to -10 dBm, with the additional contribution from nonlinear absorption of the graphene-silicon cavity membrane.

**Supplementary References:**